\documentstyle[11pt,newpasp,twoside,psfig]{article}
\def\edcomment#1{\iffalse\marginpar{\raggedright\sl#1\/}\else\relax\fi}
\reversemarginpar

\begin{document}

\title{The double pulsar -- A new testbed
for relativistic gravity}

\author{M. Kramer,$^{1}$ A.G. Lyne,$^{1}$ M. Burgay,$^{2}$ 
A. Possenti,$^{3,4}$ 
R.N.~Manchester,$^{5}$ F. Camilo,$^{6}$ M.A. McLaughlin,$^{1}$ D.R. Lorimer,$^{1}$
N.~D'Amico,$^{4,7}$ B.C. Joshi,$^{8}$ J. Reynolds$^{5}$ and P.C.C. Freire$^{9}$}
\affil{$^{1}$University of Manchester, Jodrell Bank Observatory, UK,
$^{2}$Universita degli Studi di Bologna, Dipartimento di Astronomia, 
 Italy,
$^{3}$INAF - Osservatorio Astronomica di Bologna, Italy,
$^{4}$INAF - Osservatorio Astronomica di Cagliari, Italy,
$^{5}$ATNF, CSIRO, Australia,
$^{6}$Columbia Astrophysics Laboratory, Columbia University, USA,
$^{7}$Universita degli Studi di Cagliari, Dipartimento di Fisica, Italy,
$^{8}$NCRA, Pune, India,
$^{9}$NAIC, Arecibo Observatory, USA}

\begin{abstract}
The first ever double pulsar, discovered by our team a few months ago,
consists of two pulsars, one with period of 22 ms and the other
with a period of 2.7 s. This binary system with a period of only
2.4-hr provides a truly unique laboratory for relativistic gravitational 
physics. In this contribution we summarize the published results and look
at the prospects of future observations.
\end{abstract}

\section{Introduction}

In these proceedings, the contribution by Burgay et al.~describes the
discovery of the compact double neutron star system J0737$-$3039 and
its implication for the detection rate for ground-based gravitational
wave detectors and tests of general relativity (see also Burgay et
al.~2003). 
With the discovery of radio pulses from the companion, the
system becomes unique in its potential as a testbed for relativistic
gravity. The data resulting in the timing parameters presented for
both pulsars by Lyne et al.~(2004) and shown in Table 1, were obtained
with the Parkes radio telescope and the Lovell telescope at Jodrell
Bank.  An updated timing solution covering a full year of observations
will be presented in a forthcoming paper (Kramer et al.~in prep.).
The contribution by Manchester et al.~discusses 
the implications for studying pulsar magnetospheres.

\section{Tests of general relativity}

Since neutron stars are very compact massive objects, double neutron
star (DNS) binaries can be considered as almost ideal point sources
for testing theories of gravity in the strong-gravitational-field
limit. Tests can be performed when a number of relativistic
corrections to the Keplerian description of the orbit, the so-called
``post-Keplerian'' (PK) parameters, can be measured. For point masses
with negligible spin contributions, the PK parameters in each theory
should only be functions of the a priori unknown neutron star masses,
$m_A$ and $m_B$,
and the well measurable Keplerian parameters. With the two masses as
the only free parameters, the measurement of three or more PK
parameters over-constrains the system, and thereby provides a test
ground for theories of gravity.  In a theory that describes the binary
system correctly, the PK parameters produce theory-dependent lines in
a mass-mass diagram that all intersect in a single point. (See
contributions by Weisberg \& Taylor for an update on the tests
provided by PSR B1913+16 and by Stairs et al.~for tests possible with
PSR~B1534+12.)  As shown in Table~1, we have measured PSR J0737$-$3037A's
advance of periastron, $\dot{\omega}$, and the gravitational redshift/time
dilation parameter $\gamma$ and have also detected the Shapiro delay
in the pulse arrival times of A due to the gravitational field of 
PSR J0737$-$3039B
(see Fig.~1), providing a precise measurement of the orbital
inclination of $s\equiv \sin i = 0.9995(^{+4}_{-32})$
and the ``range'' parameter $r$. This already provides
four measured PK parameters, resulting in a $m_A-m_B$ plot shown in
Figure~2 where we label $\dot{\omega}(m_A,m_B)$, $\gamma(m_A,m_B)$,
$r(m_A,m_B)$, $s(m_A,m_B)$ as predicted by general relativity (Damour
\& Deruelle 1986). 

 In addition  to tests
with these PK parameters, the detection of B as a pulsar opens up
opportunities that go well beyond what is possible with previously known
DNSs. With a measurement of the projected semi-major axes of the
orbits of both A and B (see Table 1), we obtain a precise measurement
of the mass ratio, $R(m_{\rm{A}}, m_{\rm{B}}) \equiv
m_{\rm{A}}/m_{\rm{B}} = x_{\rm{B}}/x_{\rm{A}}$, providing a further
constraint displayed in Figure~2.  For every realistic theory of
gravity, we can expect the mass ratio, $R$, to follow this simple
relation (Damour \& Taylor 1992). Most importantly, the
$R$-line is not only theory-independent, but also independent of
strong-field (self-field) effects which is not the case for
PK-parameters.  This provides a stringent and new constraint for tests
of gravitational theories as any intersection of the PK-parameters
{\em must} be located on the $R$-line in Fig.~2.
With four PK parameters already available, this additional constraint
makes J0737--3039 the most overdetermined DNS system to date where the
most relativistic effects can be studied in the strong-field limit.

\begin{table}[h]
\caption{Observed and derived parameters of PSRs~J0737$-$3039A and B.
Standard errors are given in parentheses after the values and are in
units of the least significant digit(s).}

\begin{center}
\footnotesize
\begin{tabular}{lcc}
\hline
\hline
 & & \\ 
Pulsar & PSR~J0737$-$3039A & PSR~J0737$-$3039B \\
Pulse period $P$ (ms) & 22.69937855615(6) & 2773.4607474(4) \\
Period derivative $\dot{P}$ & $1.74(5) \times 10^{-18}$ & $0.88(13)\times 10^{-15}$ \\
Epoch of period (MJD) & 52870.0 & 52870.0 \\
Right ascension (J2000) & $07^{\rm{h}}37^{\rm{m}}51^{\rm{s}}.247(2)$ & $-$ \\
Declination (J2000) & $-30^\circ 39' 40''.74(3)$ & $-$ \\
Orbital period $P_{\rm{b}}$ (day) & 0.102251563(1) & $-$ \\
Eccentricity $e$ & 0.087779(5) & $-$ \\
Epoch of periastron $T_0$ (MJD) & 52870.0120589(6) & $-$ \\
Longitude of periastron $\omega$ (deg) & 73.805(3) & 73.805 + 180.0 \\
Projected semi-major axis $x=a \sin i/c$ (sec) & 1.41504(2) & 1.513(4) \\
Advance of periastron $\dot{\omega}$ (deg/yr) & 16.90(1) & $-$ \\
Gravitational redshift parameter $\gamma$ (ms) & 0.38(5) &  \\
Shapiro delay parameter $s$ & $0.9995(-32,+4)$ & \\
Shapiro delay parameter $r$ ($\mu$s) & $5.6(-12,+18)$ &  \\
Total system mass $m_{\rm{A}}+m_{\rm{B}}$ (M$_\odot$) & \multicolumn{2}{c}{2.588(3) } \\
Mass ratio $R \equiv m_{\rm{A}}/m_{\rm{B}}$ & \multicolumn{2}{c}{1.069(6) } \\
Orbital inclination from Shapiro $s$ (deg) & \multicolumn{2}{c}{87(3)} \\
Orbital inclination from $(R,\dot\omega)$ (deg) & \multicolumn{2}{c}{$87.7(-29,+17)$} \\
Stellar mass from $(R,\dot\omega)$ (M$_\odot$) & 1.337(5) & 1.250(5) \\
 & & \\
\hline
\end{tabular}
\end{center}
\end{table}

\section{Future potential}

We expect to measure the orbital period derivative, $\dot{P}_{\rm b}$,
by autumn 2004.  We note that the distance to this
pulsar will be needed to correctly account for relative
accelerations and hence to interpret $\dot{P}_{\rm b}$ in the context of
gravitational theories. VLBI observations to determine the system's
parallax are underway. In the longer term, further effects should be
measurable.

\subsection{Geodetic Precession \& Aberration}

In general relativity, the proper reference frame of a freely-falling
object suffers a precession with respect to a distant observer, called
geodetic precession. In a binary pulsar system this geodetic
precession leads to a relativistic spin-orbit coupling, analogous of
spin-orbit coupling in atomic physics (Damour \& Ruffini 1974).
As a consequence, the pulsar spin precesses about the
total angular momentum, changing the relative orientation of the
pulsar towards Earth. Since the orbital angular momentum is much
larger than the pulsars' spins, the total angular momentum is 
practically represented by the orbital spin. The precession rate
(Barker \& O'Connell 1975, B\"o{}rner et al.~1975) depends on the
period and the eccentricity of the orbit as well as the pulsar and
companion mass.  With the parameters shown in Table 1, general
relativity predicts precession periods of only 75~yr for A and 71~yr
for B.

Geodetic precession should have a direct effect on the timing as it
causes the polar angles of the spins and hence the effects of
aberration to change with time (Damour \& Taylor 1992). 
These changes modify the {\em observed} orbital parameters, like
projected semi-major axis and eccentricity, which differ from the {\em
intrinsic} values by an aberration dependent term, potentially allowing
us to infer the system geometry.  Other consequences of geodetic
precession can be expected to be detected much sooner. These arise
from changes in the pulse shape and its polarisation properties due to
changing cuts through the emission beam as the pulsar spin axis
precesses. Due to these effects, which may complicate the TOA
determination, geodetic precession is detected for PSRs B1913+16
(Kramer 1998; Weisberg \& Taylor, these proceedings)
and B1534+12 (Arzoumanian 1995; Stairs et al., these proceedings).
For PSR B1913+16 it is used to derive a two-dimensional
map of the emission beam (Weisberg \& Taylor 2002) and to
determine the full system geometry leading to the prediction that the
pulsar will not be observable from Earth around 2025 (Kramer
1998). Since the precession rates for A and B are about
four times larger, the visible effects in J0737$-$3039 should be much
easier to detect.

\begin{figure}[hbt]

\centerline{\psfig{file=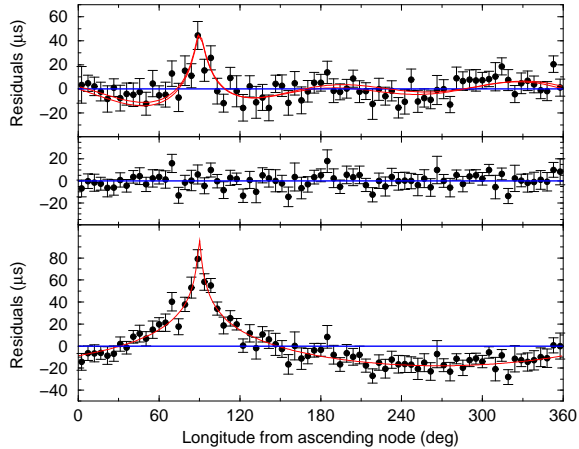,height=6cm}}

\caption{ The effect of the Shapiro delay caused by the gravitational
potential of B seen in the timing residuals of A.  Residuals were
averaged into 75 equal bins of orbital phase.  Top: timing residuals
obtained by fitting all model parameters shown in Table 1 except the
Shapiro delay parameters $r$ and $s$.  The left-over structure
represents the higher harmonics of the Shapiro delay that are
unabsorbed by fits to the Roemer delay.  Middle: Timing residuals
after fitting additionally for the Shapiro delay. Bottom: as middle
display, but with the Shapiro delay parameters $r$ and $s$ set to
zero.  }
\end{figure}

\subsection{Higher Post-Newtonian orders}

In all previous tests of general relativity involving PK parameters,
it was sufficient to compare the observed values to theoretical
expressions computed to their lowest post-Newtonian
approximation. However, higher-order corrections may become important
if relativistic effects are large and timing precision is sufficiently
high. Whilst this has not been the case in the past, the double pulsar
system may allow measurements of these effects in the future (see Lyne et
al.~2004).

\begin{figure}[hbt]

\centerline{\psfig{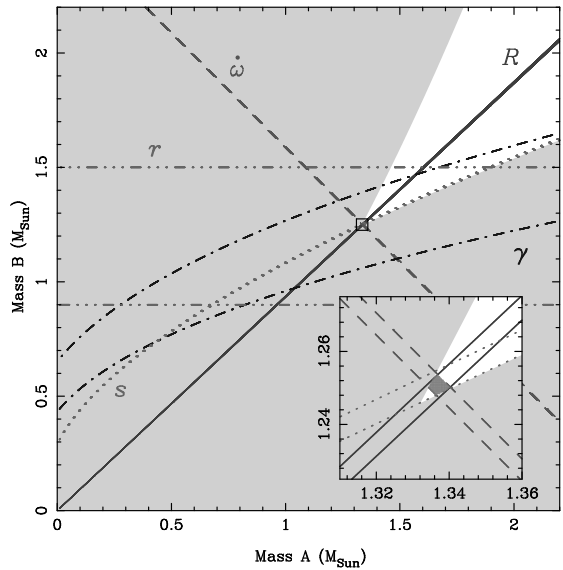}}

\caption{ The observational constraints upon the masses $m_{\rm{A}}$
and $m_{\rm{B}}$.  The solid regions are those which are excluded by
the Keplerian mass functions of the two pulsars.  Four further
constraints are shown: (a) the measurement of the advance of periastron
$\dot{\omega}$; (b) the measurement of $R=m_{\rm{B}}/m_{\rm{A}}$; (c)
gravitational redshift parameter $\gamma$ and (d) Shapiro delay
parameters $r$ and $s$.  the intersection of the four constraints,
with the scales increased by a factor of 50.  The permitted regions
are those between the pairs of parallel lines and the solid
region is the only area permitted by all constraints.  }
\end{figure}

\subsubsection{Light propagation effects}

Higher-order corrections to the present description of the 
light propagation would account for the assumptions currently
made in the computation of the Shapiro delay that gravitational
potentials are static and weak everywhere. Corresponding calculations
taking these next higher order corrections into account have been
presented by Wex (1995a),
Kopeikin \& Sch\"afer (1999) and Kopeikin
(2003).

In addition to light propagation effects causing time delays, others
exist that are related to light bending and its consequences. Light
bending as discussed by Doroshenko \& Kopeikin (1995)
would be superposed on the Shapiro delay as a typically much weaker
signal, which arises due to a modulation of the pulsars' rotational
phase by the effect of gravitational deflection of the light in the
field of the pulsar's companion. This aberration-like effect would
only be visible for very nearly edge-on systems since it depends
crucially on the orientation of the pulsar's spin-axis in space. The
double pulsar system fulfils these requirements.

\subsubsection{Moment of inertia}

In contrast to Newtonian physics, general relativity predicts that the
neutron stars' spins affect their orbital motion via spin-orbit
coupling. This effect would be most clearly visible
as a contribution to
the observed $\dot{\omega}$ in a secular (Barker \& O'Connell
1975), and periodic fashion (Wex 1995b).
For the J0737$-$3039 system, the expected contribution is about an
order of magnitude larger than for PSR B1913+16 (Lyne et al.~2004),
i.e.~$2\times 10^{-4}$ deg yr$^{-1}$ (for A, assuming a geometry as
determined for PSR B1913+16; Kramer 1998). As the exact value depends
on the pulsars' moment of inertia, a potential measurement of this
effect would allow us 
the moment of inertia of a neutron star to be determined
for the first time (Damour \& Sch\"afer 1988).
If two parameters, e.g.~the Shapiro parameter $s$ and the mass ratio
$R$, can be measured sufficiently accurately, an expected
$\dot{\omega}_{\rm exp}$ can be computed from the intersection
point. This value can be compared to the observed value
$\dot{\omega}_{\rm obs}$ which is given by (see Damour \& Sch\"afer
1988) 
\begin{equation}
\dot{\omega}_{\rm obs} = \dot{\omega}_{\rm 1PN}\;\left[
1+ \Delta\dot{\omega}_{\rm 2PN} 
  - g^A \Delta \dot{\omega}_{\rm SO}^A
  - g^B \Delta \dot{\omega}_{\rm SO}^B
\right],
\end{equation}
where the last two terms represent contributions from the pulsars' spins.
In these terms, $g^{A,B}$ are geometry-dependent factors whilst
$\Delta \dot{\omega}_{\rm SO}^{A,B}$ arise from relativistic
spin-orbit coupling (Barker \& O'Connell 1975), formally at the 
level of first post-Newtonian (1PN) approximation.  However, it turns out
that for binary pulsars these effects have a magnitude equivalent to
second post-Newtonian (2PN) order (Wex 1995b), so that they only need to be
considered if $\dot{\omega}$ is to be studied at this higher level of
approximation. We find $\Delta \dot{\omega}_{\rm SO}\propto I/P m^2$
(Damour \& Sch\"afer 1988), so that with precisely measured masses $m$
the moment of inertia $I$ can be measured and the neutron star
equation-of-state and our understanding of matter at extreme
pressure and densities can be tested.

The dependence of $\Delta \dot{\omega}_{\rm SO}$ on the spin period
$P$ suggests that only a measurement for pulsar $A$ can be
obtained. It also requires that at least two other parameters can
be measured to a similar accuracy as $\dot{\omega}$. Despite being a
tough challenge, e.g.~due to the expected profile variation caused by
geodetic precession, the prospects are promising. Simulations indicate
that a few years of high precision timing are sufficient (Kramer et
al.~in prep.).

\section{Conclusions}

The double pulsar system offers improved but also new tests
of general relativity. The current data indicate an agreement
of the observed with the expected Shapiro parameter of 
$s_{\rm obs}/s_{\rm exp} = 1.0001\pm0.00220$ (Kramer et al.~in
prep.) where the uncertainties are likely to decrease, enabling
the other discussed applications.

\acknowledgements
MK thanks the Aspen Center for Physics and the organisers for the
received financial support.

\end{document}